\documentclass[aps,prd,nofootinbib,twocolumn,preprintnumbers,floatfix,superscriptaddress,bibliography]{revtex4-1}
\usepackage[colorlinks, linkcolor=red, anchorcolor=blue, citecolor=blue, colorlinks=true, urlcolor=blue]{hyperref}
\usepackage{xspace}
\usepackage{threeparttable}
\usepackage{gensymb}
\usepackage{indentfirst}
\usepackage{subfigure}
\usepackage{ulem}
\usepackage{bm}
\usepackage{fontawesome}
\usepackage{slashed}
\usepackage{lipsum}
\usepackage[figuresright]{rotating}
\usepackage{amssymb}
\usepackage{amsmath}
\usepackage{amsthm}
\usepackage{multirow}
\usepackage{epstopdf}
\usepackage{breakurl}
\usepackage{mathrsfs}
\usepackage{graphicx}
\usepackage{cancel}
\makeatother
\allowdisplaybreaks
 
\begin{document}
 
\title{Supermassive black holes triggered by QCD axion bubbles}
 
\preprint{BNU-23-043} 

\author{Hai-Jun Li} 
\email{lihaijun@itp.ac.cn}
\affiliation{Key Laboratory of Theoretical Physics, Institute of Theoretical Physics, Chinese Academy of Sciences, Beijing 100190, China}
\affiliation{Center for Advanced Quantum Studies, Department of Physics, Beijing Normal University, Beijing 100875, China}

\author{Ying-Quan Peng}
\email{yqpenghep@mail.bnu.edu.cn}
\affiliation{Center for Advanced Quantum Studies, Department of Physics, Beijing Normal University, Beijing 100875, China}

\author{Wei Chao} 
\email{chaowei@bnu.edu.cn}
\affiliation{Center for Advanced Quantum Studies, Department of Physics, Beijing Normal University, Beijing 100875, China}

\author{Yu-Feng Zhou}
\email{yfzhou@itp.ac.cn}
\affiliation{Key Laboratory of Theoretical Physics, Institute of Theoretical Physics, Chinese Academy of Sciences, Beijing 100190, China}
\affiliation{School of Physical Sciences, University of Chinese Academy of Sciences, Beijing 100049, China}
\affiliation{School of Fundamental Physics and Mathematical Sciences, Hangzhou Institute for Advanced Study, UCAS, Hangzhou 310024, China}
\affiliation{International Centre for Theoretical Physics Asia-Pacific, Beijing/Hangzhou, China}

\date{\today}

\begin{abstract}

The supermassive black holes (SMBHs) are ubiquitous in the center of galaxies, although the origin of their massive seeds is still unknown.
In this paper, we investigate the SMBHs formation from the QCD axion bubbles.
In this case, the primordial black holes (PBHs) are considered as the seeds of SMBHs, which are generated from the QCD axion bubbles due to an explicit Peccei-Quinn (PQ) symmetry breaking after inflation.
The QCD axion bubbles are formed when the QCD axion starts to oscillate during the QCD phase transition.
We consider a general case in which the axion bubbles are formed with the bubble effective angle $\theta_{\rm eff}\in(0, \, \pi]$, leading to the minimum PBH mass $\sim\mathcal{O}(10^4-10^7)M_\odot$ for the axion decay constant $f_a\sim\mathcal{O}(10^{16})\, \rm GeV$.
The PBHs at this mass region may account for the seeds of SMBHs.
 

\end{abstract}
\maketitle

\section{Introduction}

The supermassive black holes (SMBHs) with the mass $\sim\mathcal{O}(10^6-10^9)M_\odot$ are ubiquitous in the center of galaxies \cite{Kormendy:1995er, Richstone:1998ky, Kormendy:2013dxa}, where $M_\odot$ is the solar mass.
They are also considered as the central engines of active galactic nuclei (AGN) and quasars \cite{Lynden-Bell:1969gsv, Ferrarese:2000se}.
However, the origin of SMBHs is not clear.
They may be formed from the stellar black holes (BHs) with the mass $\sim\mathcal{O}(10)M_\odot$ through accretion and mergers \cite{Page:1974he, Buonanno:2007sv}.
Nevertheless, it is difficult to account for the SMBHs at the high redshift $z\sim7$ \cite{Mortlock:2011va, DeRosa:2013iia, Banados:2017unc} since the stellar BHs have no time to grow into SMBHs.
The another scenario is considering the primordial black holes (PBHs) with the mass $\sim\mathcal{O}(10^4-10^5)M_\odot$ as the seeds of SMBHs \cite{Salpeter:1964kb, Duechting:2004dk}.
In this case, the PBHs could subsequently grow up to $\sim\mathcal{O}(10^9)M_\odot$ due to an efficient accretion of matter on the massive seeds and mergings \cite{Choquette:2018lvq, Dolgov:2019vlq}.
Additionally, the SMBHs may also have the origin from the collapse of self-interaction dark matter (DM) \cite{Spergel:1999mh, Kaplinghat:2015aga, Feng:2020kxv, Feng:2021rst}.
See $\rm e.g.$ Refs.~\cite{Volonteri:2010wz, Dolgov:2017aec} for reviews of the SMBHs formation.

The PBHs have recently gathered significant interest as the candidates of DM \cite{Bird:2016dcv, Green:2020jor, Carr:2020xqk, Carr:2020gox, Carr:2021bzv, Gelmini:2022nim, Agashe:2022jgk, Li:2022mcf, Jho:2022wxd, Agashe:2022phd, Papanikolaou:2023nkx, Xie:2023cwi, Gelmini:2023ngs}.
They can be formed due to the collapse of large density fluctuations produced during inflation \cite{Carr:1975qj, Ivanov:1994pa, Garcia-Bellido:1996mdl, Alabidi:2009bk, Ashoorioon:2020hln}, the cosmological phase transitions \cite{Kodama:1982sf, Hawking:1982ga, Baker:2021nyl}, and the collapse of topological defects \cite{Hawking:1987bn, Polnarev:1988dh, MacGibbon:1997pu, Rubin:2001yw, Deng:2016vzb, Ferrer:2018uiu, Ge:2019ihf} and false vacuum bubbles \cite{Deng:2017uwc, Kusenko:2020pcg, Deng:2020mds, Maeso:2021xvl}, etc.
The another scenario is considering the PBHs formation from the QCD axion bubbles in the early Universe, leading to the minimum PBH mass $\sim\mathcal{O}(10^4)M_\odot$ \cite{Kitajima:2020kig}.  
This scenario is similar to the PBHs formation from the baryon bubbles in the Affleck-Dine baryogenesis \cite{Dolgov:2008wu, Hasegawa:2018yuy}.
The QCD axion bubbles can be generated due to an explicit PQ symmetry breaking after inflation with the multiple vacua, and are formed when the QCD axion starts to oscillate during the QCD phase transition.
The axion acquires a light mass after inflation, then starts to oscillate when this mass is comparable to the Hubble parameter and settles down into different potential minima, depending on its initial position.
Since the explicit PQ symmetry is supposed to be temporarily broken, the axion potential will disappear before the QCD phase transition.
Therefore, the misalignment mechanism is still valid to calculate the final QCD axion abundance.
However, the initial misalignment angle will be split into different values by the multiple vacua.
During the QCD phase transition, the QCD axions start to oscillate with these different initial angles, in which the high density axion bubbles can be produced with the large initial values.
An interesting phenomenon of the QCD axion bubbles is the PBHs formation, which can be formed when the axions dominate the radiation in the bubbles.

In this paper, we focus our attention on the PBHs triggered by QCD axion bubbles as the seeds of SMBHs.
The axion acquires a light mass due to an explicit PQ symmetry breaking after inflation with a large axion decay constant $f_a\sim\mathcal{O}(10^{16})\, \rm GeV$, and then settles down into two potential minima, $\phi_{\rm min}^0$ and $\phi_{\rm min}^1$.
We consider a general case in which the QCD axion bubbles are formed with the bubble effective angle $\theta_{\rm eff}\in(0, \, \pi]$.
The $\phi_{\rm min}^0$ accounts for the DM abundance with the initial misalignment angle $\sim0$, while the $\phi_{\rm min}^1$ forms the high energy density QCD axion bubbles with the angle $\sim\theta_{\rm eff}$.
The PBHs triggered by QCD axion bubbles are formed after the QCD phase transition, which leads to the minimum PBH mass $\sim\mathcal{O}(10^4-10^7)M_\odot$ for $f_a\sim\mathcal{O}(10^{16})\, \rm GeV$.
Compared with the critical value of the $\sim\mathcal{O}(10^9)M_\odot$ SMBHs seeds, the PBHs at this mass region may account for the seeds of SMBHs.

The paper is structured as follows.
In Sec.~\ref{sec_QCD axion bubbles}, we introduce the QCD axion bubbles scenario.
In Sec.~\ref{sec_SMBHs}, we investigate the PBHs triggered by QCD axion bubbles as the seeds of SMBHs.
The conclusion is given in Sec.~\ref{sec_conclusion}.

\section{QCD axion bubbles with $\theta_{\rm eff}$}
\label{sec_QCD axion bubbles} 

In this section, we introduce the QCD axion bubbles scenario \cite{Kitajima:2020kig}.
The effective potential of the QCD axion is given by
\begin{eqnarray}
V_{\rm QCD}(\phi)=m_a^2(T) f_a^2\left[1-\cos\left(\frac{\phi}{f_a}\right)\right]\, ,
\label{eq_Va}
\end{eqnarray}
with the axion field $\phi$, the axion decay constant $f_a$, the axion angle $\theta=\phi/f_a$, and the temperature-dependent axion mass
\begin{eqnarray}
\begin{aligned}
m_a(T)\simeq
\begin{cases}
m_{a,0}\left(\dfrac{T}{T_{\rm QCD}}\right)^{-4.08}\, , & T\geq T_{\rm QCD} \\
m_{a,0}\, , & T< T_{\rm QCD} 
\end{cases} 
\end{aligned}
\end{eqnarray} 
where $m_{a,0}\simeq 5.70(7)\,{\mu \rm eV}(10^{12}\,{\rm GeV}/f_a)$ is the zero-temperature axion mass \cite{GrillidiCortona:2015jxo}, $T$ is the cosmic temperature, and $T_{\rm QCD}\simeq150\, \rm MeV$.

During the QCD phase transition, the QCD axion starts to oscillate when its mass $m_a(T)$ is comparable to the Hubble parameter $H(T)$ at the oscillation temperature $T_a$.
The axion energy density at present can be described by
\begin{eqnarray}
\rho_a(T_0)=\frac{m_{a,0} m_a(T_a) s(T_0)}{2s(T_a)} f_a^2\left\langle\theta_i^2f(\theta_i)\right\rangle\chi \, ,
\end{eqnarray}
with the entropy density $s(T)$, the present CMB temperature $T_0$, the initial misalignment angle $\theta_i$, the numerical factor $\chi\simeq1.44$, and the anharmonic factor $f(\theta_i)$ is given by \cite{Lyth:1991ub}
\begin{eqnarray}
f(\theta_i)\simeq\left[\ln\left(\frac{e}{1-\theta_i^2/\pi^2}\right)\right]^{1.16}\, .
\end{eqnarray}
Then we can derive the present QCD axion abundance
\begin{eqnarray}
\begin{aligned}
\Omega_ah^2&\simeq0.14\left(\frac{g_*(T_a)}{61.75}\right)^{-0.42}\left(\frac{g_{*s}(T_0)}{3.94}\right)\\
&\times\left(\frac{f_a}{10^{12} \rm \, GeV}\right)^{1.16}\left\langle\theta_i^2f(\theta_i)\right\rangle\, ,
\label{eq_Omega_axion} 
\end{aligned}
\end{eqnarray}
where $g_*(T)$ and $g_{*s}(T)$ are the numbers of effective degrees of freedom of the energy density and the entropy density, respectively, and $h\simeq0.68$ is the reduced Hubble constant.
In order to account for the observed cold DM abundance, the initial misalignment angle should be $\theta_i\sim\mathcal{O}(1)$ for the scale $f_a\sim\mathcal{O}(10^{12})\, \rm GeV$.

\begin{figure}[t] 
\centering
\includegraphics[width=0.48\textwidth]{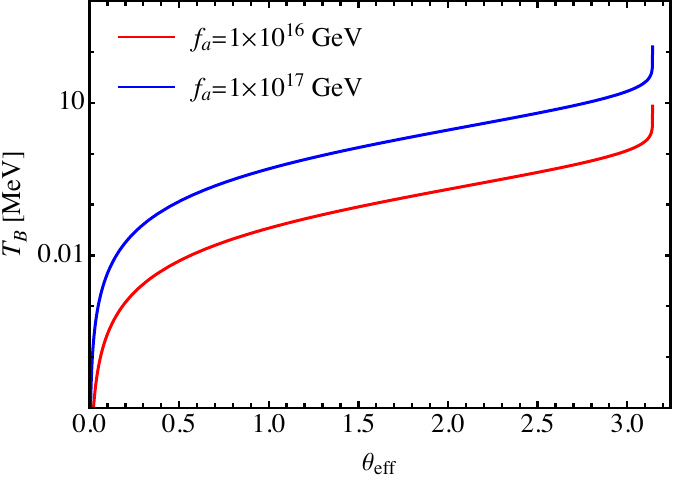}
\caption{The temperature $T_B$ as a function of the axion bubble effective angle $\theta_{\rm eff}$.
The red and blue lines represent the scales $f_a=1\times10^{16}\, \rm GeV$ and $1\times10^{17}\, \rm GeV$, respectively.
}
\label{fig_TB}
\end{figure}

The QCD axion bubbles can be generated due to an explicit PQ symmetry breaking after inflation.
There are many mechanisms for this explicit PQ symmetry breaking in the early Universe \cite{Nomura:2015xil, Kawasaki:2015lpf, Kawasaki:2017xwt, Chiba:2003vp, Takahashi:2003db, Higaki:2014ooa, Choi:1996fs, Banks:1996ea, Jeong:2013xta, Takahashi:2015waa, Li:2023det}.
The axion field acquires the effective potential, which can be approximated as
\begin{eqnarray}
V_{\cancel{\rm PQ}}(\phi)\simeq \frac{1}{2}m_a^2(\phi) \left(\phi-\phi_{\rm min}^n\right)^2\, ,
\end{eqnarray} 
where $m_a(\phi)$ is the axion effective mass, $\phi_{\rm min}^n$ is the potential minimum, and $n$ is an integer.
The explicit PQ symmetry is assumed to be temporarily broken with a large scale $f_a\sim\mathcal{O}(10^{16})\, \rm GeV$.
The axion acquires quantum fluctuations and settles down into the minimum $\phi_{\rm min}^n$.
The QCD axion bubbles are generated when the conventional axion potential $V_{\rm QCD}(\phi)$ arises during the QCD phase transition.
Therefore, one of the minimum $\phi_{\rm min}^0$ of $V_{\cancel{\rm PQ}}(\phi)$ should be near the minimum of $V_{\rm QCD}(\phi)$, ensuring the correct cold DM abundance with the effective initial angle $\theta_{i,0}$.
In addition, if the initial value is greater than a critical value $\phi_{\rm crit}$, the axion will be stabilized at the another minimum $\phi_{\rm min}^1$ with $\theta_{i,1}$.
Here we suppose that the QCD axion bubbles are produced at $\phi_{\rm min}^1$, $\rm i.e.$, $\theta_{i,1}>\theta_{i,0}$, we define the axion bubble effective angle $\theta_{\rm eff}\in(0, \, \pi]$.
Then the effective initial misalignment angle are given by
\begin{eqnarray}
\theta_{i,n}=
\begin{cases}
0-\theta_i \, , & n=0 \\
\theta_{\rm eff}-\theta_i \, , & n=1 
\end{cases} 
\label{theta_i_n}
\end{eqnarray} 
corresponding to the potential minima $\phi_{\rm min}^0$ and $\phi_{\rm min}^1$, respectively.
Note that $\theta_i$ in Eq.~(\ref{theta_i_n}) is a small initial misalignment angle 
\begin{eqnarray}
\begin{aligned}
\theta_i&\simeq4.29\times10^{-3}\left(\frac{g_*(T_a)}{61.75}\right)^{0.21}\left(\frac{g_{*s}(T_0)}{3.94}\right)^{-1/2}\\
&\times\left(\frac{f_a}{10^{16}\, \rm GeV}\right)^{-0.58}\, .
\label{eq_theta_i} 
\end{aligned}
\end{eqnarray}
When the axions dominate the radiation in the bubbles, the cosmic background temperature is defined as $T_B$ \cite{Kitajima:2020kig}.
The axion energy density at $T_B$ is equal to the radiation energy density, $\rho_a(T_B)=\rho_R(T_B)$, where $\rho_R$ is the radiation energy density.
Since the axion energy density inside the bubbles at $T_B$ can be described by
\begin{eqnarray}
\begin{aligned}
\rho_a(T_B)&=\frac{m_{a,0} m_a(T_a) s(T_B)}{2s(T_a)}\\
&\times f_a^2\left\langle\left(\theta_{\rm eff}-\theta_i\right)^2f(\theta_{\rm eff}-\theta_i)\right\rangle\chi \, ,
\end{aligned}
\end{eqnarray}
then we can derive the temperature $T_B$
\begin{eqnarray}
\begin{aligned}
T_B&\simeq2.13\times10^{-2}\, {\rm MeV}\\
&\times\left(\frac{g_*(T_a)}{61.75}\right)^{-0.42}\left(\frac{f_a}{10^{16}\, \rm GeV}\right)^{1.16}\\
&\times\left\langle\left(\theta_{\rm eff}-\theta_i\right)^2f(\theta_{\rm eff}-\theta_i)\right\rangle\chi \, .
\label{eq_TB}
\end{aligned}
\end{eqnarray}
Using Eq.~(\ref{eq_theta_i}), we show $T_B$ as a function of the effective angle $\theta_{\rm eff}$ in Fig.~\ref{fig_TB}.
Two typical values of the scales $f_a=1\times10^{16}\, \rm GeV$ (red) and $1\times10^{17}\, \rm GeV$ (blue) are selected for comparisons. 
We find that the temperature $T_B\sim\mathcal{O}(0.1-1)\,\rm MeV$ decreases slowly as the effective angle $\theta_{\rm eff}$ decreases, and varies most rapidly when the angle approaches a small value.

The QCD axion bubbles abundance is related to the inflationary fluctuations.
The volume fraction of the bubbles can be described by \cite{Kitajima:2020kig}
\begin{eqnarray}
\frac{{\rm d}\beta}{{\rm d}\ln k}\simeq\frac{\phi_{\rm crit}-\phi_i}{2}P(\ln(k/k_*),\phi_{\rm crit})\,,
\label{eq_volume_fraction}
\end{eqnarray} 
where $\phi_{\rm crit}$ is the critical value, $k$ is the wave number, $k_*=0.002\, \rm Mpc^{-1}$, and $P(\ln(k/k_*),\phi)$ is the probability density function
\begin{eqnarray}
\begin{aligned}
P(\ln(k/k_*),\phi)&=\frac{1}{\sqrt{2\pi}\sigma(\ln(k/k_*))}\\
&\times\exp{\left(-\frac{\left(\phi-\phi_i\right)^2}{2\sigma^2(\ln(k/k_*))}\right)}\, , 
\end{aligned}
\end{eqnarray}
where $\sigma(\ln(k/k_*))=H_{\rm inf}\sqrt{\ln(k/k_*)}/(2\pi)$ is the variance, $P(\ln(k/k_*),\phi)=\delta(\phi-\phi_i)$ is the initial condition, and $H_{\rm inf}$ is the Hubble parameter during inflation.
Here we take $\phi_{\rm crit}-\phi_i=4.5H_{\rm inf}$ as a benchmark.
 
\begin{figure}[t] 
\centering
\includegraphics[width=0.48\textwidth]{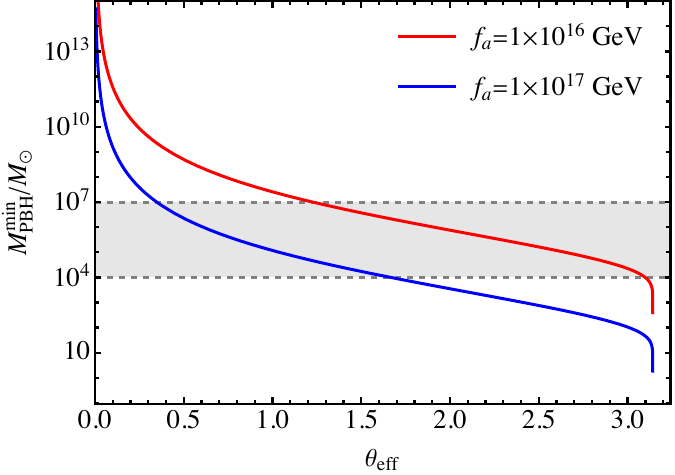}
\caption{The minimum PBH mass $M_{\rm PBH}^{\rm min}$ (in the solar mass $M_\odot$) as a function of the angle $\theta_{\rm eff}$.
The red and blue lines represent $f_a=1\times10^{16}\, \rm GeV$ and $1\times10^{17}\, \rm GeV$, respectively.
The gray shadow region represents the PBH mass region $\sim\mathcal{O}(10^4-10^7)M_\odot$.
}
\label{fig_mmin}
\end{figure}

\section{SMBHs from the QCD axion bubbles}
\label{sec_SMBHs}  

The astronomical observations indicate that SMBHs are ubiquitous in the center of galaxies with the mass $\sim10^9 M_\odot$ at the redshift $z\sim7$ ($\sim 0.76\, \rm Gyr$) \cite{Mortlock:2011va, DeRosa:2013iia, Banados:2017unc}.  
However, the origin of such BHs is still a mystery.
The mass of an accreting BH with the time $t$ is given by \cite{Salpeter:1964kb}
\begin{eqnarray}
M(t)=M_0\exp{\left(\frac{1-\epsilon_r}{\epsilon_r}\frac{t}{t_E}\right)}\, ,
\end{eqnarray}
where $M_0$ is the seed BH mass, $\epsilon_r\simeq0.1$ is the radiative efficiency, and $t_E\simeq0.45\, \rm Gyr$.
In this case, a seed BH with the mass $M_0\sim\mathcal{O}(10^2-10^5)M_\odot$ would take at least $\sim0.5\, \rm Gyr$ to grow up to a $\sim10^9 M_\odot$ SMBH \cite{Choquette:2018lvq}.
Therefore, the SMBHs must either have the heavy seeds or have a primordial origin.

\begin{figure*}[t] 
\centering
\includegraphics[width=0.48\textwidth]{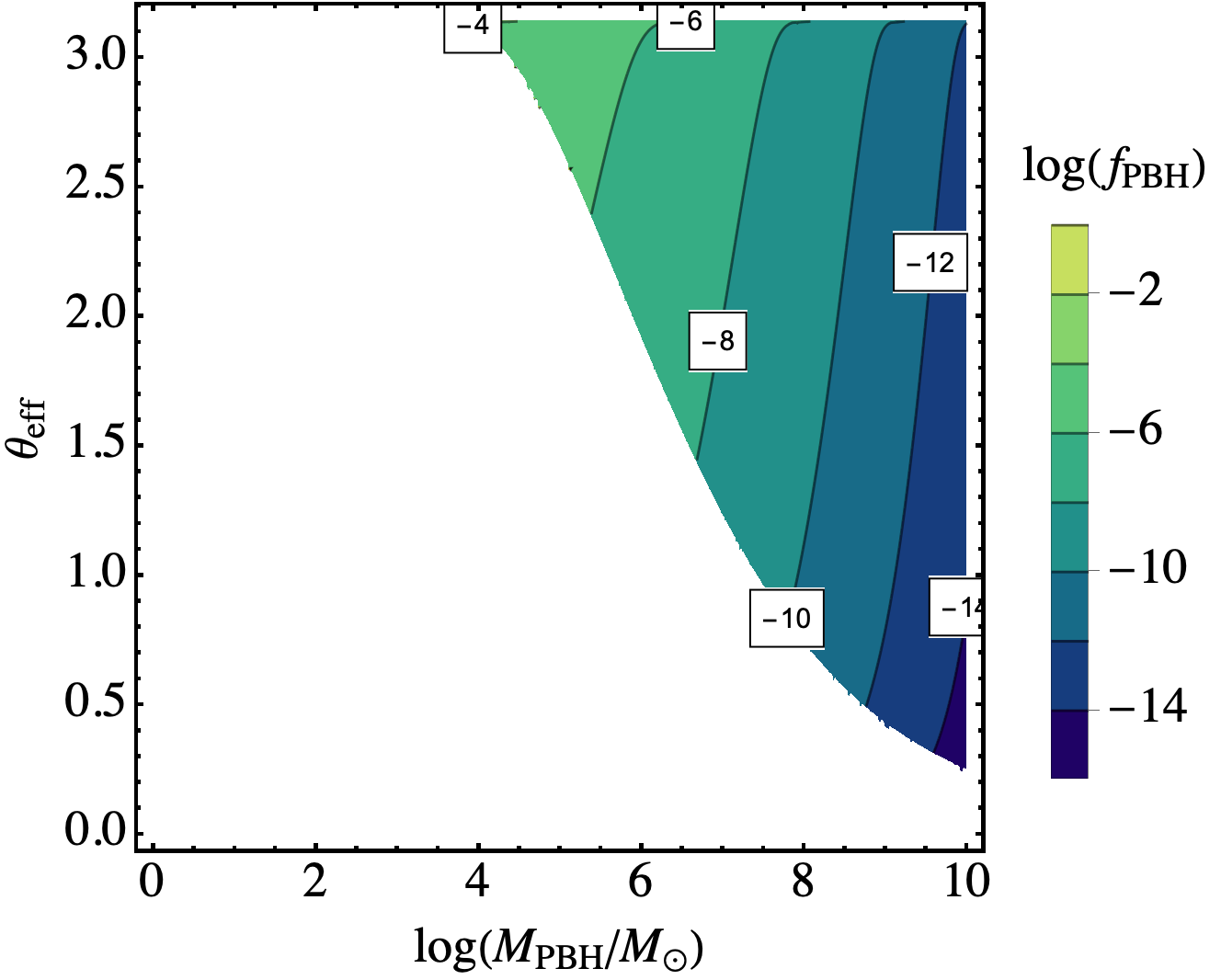}\quad\includegraphics[width=0.48\textwidth]{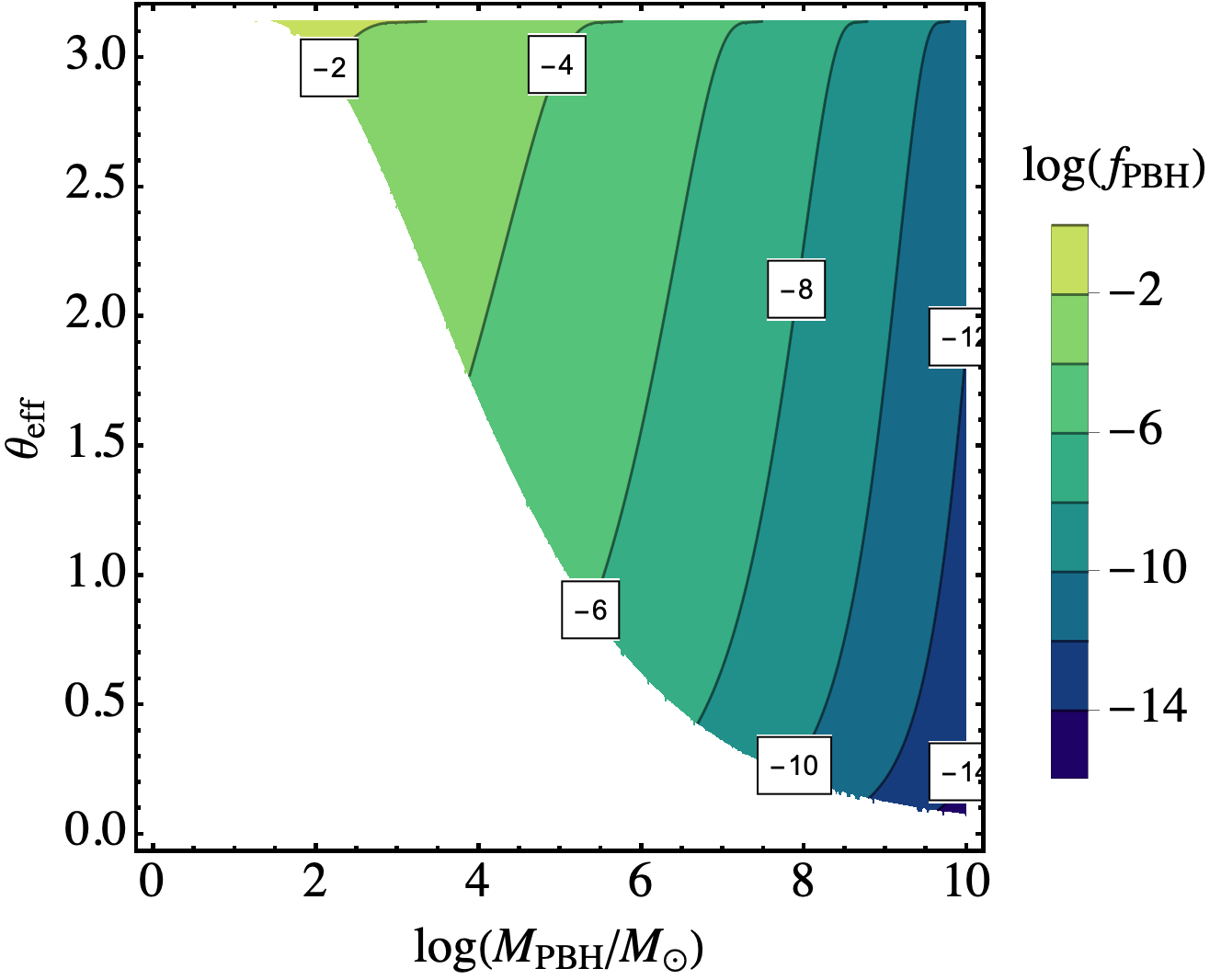}
\caption{The distributions of the PBH fractional abundance $\log(f_{\rm PBH})$ in the $\{\log(M_{\rm PBH}/M_\odot), \theta_{\rm eff}\}$ plane for $f_a=1\times10^{16}\, \rm GeV$ (left) and $1\times10^{17}\, \rm GeV$ (right).
The shadow regions represent the values of $\log(f_{\rm PBH})$.
The blank regions represent the limits set by $M_{\rm PBH}^{\rm min}$.
}
\label{fig_abundance_1}
\end{figure*}

Here we investigate the PBHs as the seeds of SMBHs, which are produced from the QCD axion bubbles after the QCD phase transition.
The initial PBH mass at the formation temperature $T_f$ is \cite{Carr:2009jm}
\begin{eqnarray}
M_{\rm PBH}=\frac{4\pi}{3}\frac{\gamma\rho_R(T_f)}{H^3(T_f)} \, , 
\label{eq_pbh_initial_mass}
\end{eqnarray}
where $\gamma\simeq0.2$ is the gravitational collapse factor \cite{Carr:1975qj}, and $H(T_f)$ is the Hubble parameter at $T_f$.
Considering the PBHs triggered by QCD axion bubbles, they will be produced when the axions dominate the radiation in the bubbles ($T_f\lesssim T_B$), and the bubble size is larger than the horizon size.
Using Eqs.~(\ref{eq_TB}) and (\ref{eq_pbh_initial_mass}), we can derive the minimum PBH mass in the axion bubbles scenario
\begin{eqnarray}
\begin{aligned}
\frac{M_{\rm PBH}^{\rm min}}{M_\odot}&\simeq6.58\times10^7\left(\frac{\gamma}{0.2}\right)\left( \frac{g_*(T_a)}{61.75}\right)^{0.84}\\
&\times\left(\frac{g_*(T_f)}{10.75}\right)^{-1/2}\left(\frac{f_a}{10^{16}\, \rm GeV}\right)^{-2.33}\\
&\times\left(\left\langle\left(\theta_{\rm eff}-\theta_i\right)^2f(\theta_{\rm eff}-\theta_i)\right\rangle\chi\right)^{-2}\, .
\label{eq_M_PBH_min}
\end{aligned} 
\end{eqnarray}
We show $M_{\rm PBH}^{\rm min}$ as a function of the bubble effective angle $\theta_{\rm eff}$ in Fig.~\ref{fig_mmin}.
The red and blue lines represent $f_a=1\times10^{16}\, \rm GeV$ and $1\times10^{17}\, \rm GeV$, respectively.
For $f_a=1\times10^{16}\, \rm GeV$, we find the minimum PBH mass is roughly at $\sim\mathcal{O}(10^4-10^7)M_\odot$, which is shown with the gray shadow region.
This mass region would be smaller for $f_a=1\times10^{17}\, \rm GeV$.
Considering the PBHs as the seeds of SMBHs, the numerical simulations show that PBHs with the mass $\sim\mathcal{O}(10^4-10^5)M_\odot$ could subsequently grow up to the $\sim\mathcal{O}(10^9)M_\odot$ SMBHs \cite{2016MNRAS.462..190R}.
In this case, the PBHs mass larger than a critical value $M_c$ are considered as the seeds of SMBHs
\begin{eqnarray}
M_c=\left(10^{4}-10^{5}\right)M_\odot \, .
\end{eqnarray}
Therefore, the seed PBH mass in the QCD axion bubbles scenario is
\begin{eqnarray}
M_s={\rm Max}\left[M_{\rm PBH}^{\rm min}, \, M_c\right] \, .
\label{eq_seed_mass}
\end{eqnarray}

Then we discuss the PBH abundance. 
The energy density of PBH at the temperature $T_B$ is given by 
\begin{eqnarray}
\rho_{\rm PBH}(T_B)=\frac{3}{4}T_B s(T_B)\frac{{\rm d}\beta}{{\rm d}\ln k}\, .
\end{eqnarray}
The fractional abundance of PBH at present can be described by
\begin{eqnarray}
f_{\rm PBH}=\frac{3}{4}T_B s(T_0)\frac{{\rm d}\beta}{{\rm d}\ln k}\frac{1}{\rho_c}\frac{1}{\Omega_{\rm DM}}\, ,
\end{eqnarray}
where $\Omega_{\rm DM}\simeq0.268$ is the total cold DM abundance, and $\rho_c$ is the critical energy density.
Using Eq.~(\ref{eq_volume_fraction}), we show the distributions of $\log(f_{\rm PBH})$ in the parameters $\{\log(M_{\rm PBH}/M_\odot), \theta_{\rm eff}\}$ plane in Fig.~\ref{fig_abundance_1}.
The left and right panels correspond to the scales $f_a=1\times10^{16}\, \rm GeV$ and $1\times10^{17}\, \rm GeV$, respectively.
For the same $\theta_{\rm eff}$, the value of $f_{\rm PBH}$ decreases as the PBH mass $M_{\rm PBH}$ increases.
Note that the blank regions in the plots represent the limits set by $M_{\rm PBH}^{\rm min}$.
For $f_a\sim\mathcal{O}(10^{16})\, \rm GeV$, the PBHs at the mass region $\sim\mathcal{O}(10^4-10^7)M_\odot$ may account for the seeds of SMBHs.
While for $f_a\sim\mathcal{O}(10^{17})\, \rm GeV$, we note that the PBH fractional abundance at the mass region $\sim\mathcal{O}(10-10^4)M_\odot$ is strongly constrained by the CMB anisotropies measured by Planck \cite{Serpico:2020ehh}.

Finally, we comment on the SMBHs formation from the baryon bubbles in the Affleck-Dine baryogenesis \cite{Kawasaki:2019iis}.
The SMBHs formation in the QCD axion bubbles scenario has some similarities with that mechanism.
They considered the high density baryon bubbles formation from the modified Affleck-Dine mechanism with the Hubble induced mass and the finite temperature effect.
The PBHs will be formed when the baryon bubbles enter the horizon with the sufficiently large densities.
They also showed the PBHs produced by that mechanism may have a reasonable mass as the seeds of SMBHs.
In addition, we note that a recent discussion about the SMBHs formation from the QCD axion in Ref.~\cite{Fukuyama:2023lxc}, they considered the attractive self-interaction of QCD axion DM to form the SMBHs.

\section{Conclusion}
\label{sec_conclusion}

In summary, we have investigated the SMBHs formation from the QCD axion bubbles.
The PBHs generated from the QCD axion bubbles are considered as the seeds of SMBHs.
The QCD axion bubbles can be generated due to an explicit PQ symmetry breaking after inflation.
In this work, we do not discuss the specific mechanism of the explicit PQ symmetry breaking.
We consider a general case in which the QCD axion bubbles are formed with the bubble effective angle $\theta_{\rm eff}$.
The PBHs triggered by axion bubbles will be formed when the axions dominate the radiation in the bubbles, which leads to a minimum PBH mass $M_{\rm PBH}^{\rm min}$.
We show the distributions of the temperature $T_B$ with different $\theta_{\rm eff}$ for the scales $f_a=1\times10^{16}\, \rm GeV$ and $1\times10^{17}\, \rm GeV$, and also the resulting minimum PBH mass.
In order to explain the SMBHs in the galactic centers with the mass $\sim10^9 M_\odot$ at $z\sim7$ with the PBH scenario, the PBH mass should be larger than a critical value $\sim\mathcal{O}(10^4-10^5)M_\odot$. 
The PBH fractional abundance distributions in the $\{\log(M_{\rm PBH}/M_\odot), \theta_{\rm eff}\}$ plane are also shown, which is strongly limited by $M_{\rm PBH}^{\rm min}$.
We find that the PBHs triggered by QCD axion bubbles with the mass $\sim\mathcal{O}(10^4-10^7)M_\odot$ for the scale $f_a\sim\mathcal{O}(10^{16})\, \rm GeV$ may account for the seeds of SMBHs.
The SMBHs formation in the QCD axion bubbles scenario has some similarities with that in the baryon bubbles scenario from the modified Affleck-Dine baryogenesis.
   

{\bf Acknowledgments.}
W.C. is supported by the National Natural Science Foundation of China (NSFC) (Grants No.~11775025 and No.~12175027).
Y.F.Z. is supported by the National Key R\&D Program of China (Grant No.~2017YFA0402204), the CAS Project for Young Scientists in Basic Research YSBR-006, and the NSFC (Grants No.~11821505, No.~11825506, and No.~12047503).

 
  
\bibliography{references}
\end{document}